\documentstyle[12pt]{article}
\newcommand{\be}{\begin{equation}}
\newcommand{\ee}{\end{equation}}
\newcommand{\bea}{\begin{eqnarray}}
\newcommand{\eea}{\end{eqnarray}}
\newcommand{\ba}{\begin{array}}
\newcommand{\ea}{\end{array}}

\def\bbox{{\,\lower0.9pt\vbox{\hrule \hbox{\vrule height 0.2 cm
\hskip 0.2 cm \vrule height 0.2 cm}\hrule}\,}}
\newcommand{\dsl}{\pa \kern-0.5em /}

\newcommand{\nn}{\nonumber \\}

\begin{document}


\begin{titlepage}
\rightline{DAMTP-2003-7}
\rightline{hep-th/0301241}

\vfill

\begin{center}
\baselineskip=16pt
{\Large\bf Odd coset quantum mechanics}
\vskip 0.3cm
{\large {\sl }}
\vskip 10.mm
{\bf ~Evgeny Ivanov$^{*,1}$, Luca Mezincescu$^{\dagger,2}$, ~{}Anatoly
Pashnev
$^{*,3}$ \\
and  ~Paul K. Townsend$^{+,4}$ } \\
\vskip 1cm
{\small
$^*$
Bogoliubov Laboratory of Theoretical Physics \\
JINR, 141980 Dubna, Russia\\
}
\vspace{6pt}
{\small
$^\dagger$
Department of Physics,\\
University of Miami,\\
Coral Gables, FL 33124, USA\\
}
\vspace{6pt}
{\small
 $^+$
DAMTP, Center for Mathematical Sciences,\\
University of Cambridge, \\
Wilberforce Road, Cambridge CB3 0WA, UK\\
}
\end{center}
\vfill

\par
\begin{center}
{\bf ABSTRACT}
\end{center}
\begin{quote}

The standard quantum states of $n$ complex Grassmann variables with a
free-particle Lagrangian transform as a spinor of $SO(2n)$. However, the
same `free-fermion' model has a non-linearly realized $SU(n|1)$ symmetry;
it can be viewed as the mechanics of a `particle' on the Grassmann-odd
coset space $SU(n|1)/U(n)$. We implement a quantization of this model for
which the states with non-zero norm transform as a representation of
$SU(n|1)$, the representation depending on the $U(1)$ charge of
the wave-function. For $n=2$ the wave-function can be interpreted
as a BRST superfield.

\vfill
 \hrule width 5.cm
\vskip 2.mm
{\small
\noindent $^1$ eivanov@thsun1.jinr.ru\\
\noindent $^2$ mezincescu@server.physics.miami.edu\\
\noindent $^3$ pashnev@thsun1.jinr.ru\\
\noindent $^4$ p.k.townsend@damtp.cam.ac.uk \\
}
\end{quote}
\end{titlepage}
\setcounter{equation}{0}
\section{Introduction}

Pseudo-classical mechanics models, with anticommuting variables, have
found various applications. One class of applications is to the
pseudo-classical description of spin. Consider the `free-fermion'
Lagrangian
\be\label{lag1}
L = i\gamma\,\bar\zeta\cdot \dot \zeta
\ee
for $n$ complex anticommuting variables $\zeta^i$ and their complex
conjugates $\bar\zeta_i$ ($\gamma$ is a real, positive, dimensionless
coupling constant).
This Lagrangian has an obvious $U(n)$ invariance
but it is also invariant under the larger group $SO(2n)$. In a (coherent
state) basis for which the quantum operators corresponding to the
variables $\bar\zeta_i$ are diagonal, with eigenvalues $\bar\zeta_i$, the
Hilbert space of the quantum theory is spanned by anti-holomorphic
functions $\Phi(\{\bar\zeta\})$. This space has dimension $2^n$ and
carries a spinor representation of $SO(2n)$.

The above Lagrangian is also invariant, although less obviously, under the
following non-linear, and non-analytic, transformations
\be\label{susy1}
\delta_\epsilon \zeta^i = \left[1+ \zeta \cdot\bar\zeta\right]^{1\over2} \
\epsilon^i +  {\left(\bar\epsilon \cdot \zeta +
\epsilon\cdot\bar\zeta\right)\over 2 \left[1+ \zeta
\cdot\bar\zeta\right]^{1\over2}}\
\zeta^i
\ee
where $\epsilon^i$ are constant anticommuting parameters.
These transformations close on those of $U(n)$ to form the
superalgebra $SU(n|1)$. In other words, the above free-fermion Lagrangian
provides a non-linear realization of the supergroup $SU(n|1)$, with
$\zeta^i$ parametrizing the Grassmann-odd coset $SU(n|1)/U(n)$.

The full symmetry group is actually much larger than either $SO(2n)$ or
$SU(n|1)$; it is the supergroup of supersymplectic diffeomorphisms of a
superspace of real dimension $(0|2n)$, which is generated by all $2^{2n}$
functions on the Grassmann-odd phase-space. An alternative
characterization of it is as the closure of its two subgroups $SO(2n)$
and $SU(n|1)$. Both subgroups contain $U(n)$, which acts in an
obvious way on quantum wave-functions, so the Hilbert space decomposes
into representations of $U(n)$. In the standard quantum theory these
representations combine to yield the spinor of $SO(2n)$ and the Hilbert
space norm is the standard scalar product of two spinors.

However, one could attempt to combine the $U(n)$ representations into
representations of $SU(n|1)$ rather than $SO(2n)$. In this case, the
`Hilbert' space would be a vector superspace of
dimension $(2^{n-1}|2^{n-1})$ rather than a vector space of dimension
$2^n$, so this quantization of the free-fermion model would be very
different from the standard one; a motivation for
considering this possibility is that the $n=2$ `Hilbert' space would
then carry a representation of the Euclidean BRST group $SU(2|1)\cong
OSp(2|2)$ (see e.g. \cite{siegel}).

We shall show here that this
non-standard quantization can be implemented, but the result depends on
the resolution of an operator ordering ambiguity which leads to an
ambiguity in the definition of the $U(1)$ charge or, equivalently, the
assignment of $U(1)$ charge to the wave-function. There is a `natural'
choice, for which the $U(1)$ charge is the direct quantum analog of the
$U(1)$ Noether charge of the Lagrangian (\ref{lag1}) but, for
completeness, we consider other choices too. In many cases the
`Hilbert' space has zero norm states so the physical states in `Hilbert'
space should be taken to be the equivalence classes of states with
non-zero norm  modulo the addition of zero norm states. The $SU(n|1)$
representation content of the physical `Hilbert' space depends on the
$U(1)$ charge assigned to the wave-function. For the `natural'
resolution of the operator ordering, and $\gamma = n-1$,
we find that the physical Hilbert space is an $SU(n|1)$ singlet!

Presumably, these results could be derived by a direct attempt to
implement the $SU(n|1)$ symmetry on the `Hilbert' space found by
canonical quantization of (\ref{lag1}) but the non-analyticity of the
non-linear transformations (\ref{susy1}) makes it difficult to see how
to
do this. We
can overcome this problem by introducing the new variables
\be
\xi^i = {\zeta^i\over \left[1+ \zeta \cdot\bar\zeta\right]^{1\over2}}\,,
\ee
in terms of which the $SU(n|1)$ supersymmetry transformations are
analytic:
\be
\label{antransf}
\delta_\epsilon \xi^i = \epsilon^i + \bar \epsilon\cdot \xi\, \xi^i\,.
\ee
The Lagrangian in these new variables is\footnote{There is some similarity
to the QM reduction of the Volkov-Akulov model \cite{VA} which realizes
Poincar\'e supersymmetry non-linearly in terms of a Goldstino
variable.}
\be
\label{nonlinlag}
L =  i\,\gamma \left[1 + \bar\xi\cdot \xi\right]^{-1} \bar\xi\cdot \dot
\xi\,.
\ee
This Lagrangian can be shown to be the 1-dimensional pullback of
the $U(1)$ connection one-form in the nonlinear realization of $SU(n|1)$ in
the Grassmann-odd coset space $SU(n|1)/U(n)$. It is shifted by a total
derivative
under the $SU(n|1)$ transformations and so is a sort of 1-dimensional
Wess-Zumino (WZ) term. It contains `interactions' which complicate
the canonical quantization
procedure, but this problem is easily solved in a way that will
now be described.

\section{Analytic quantization}

An equivalent phase-space form of the  Lagrangian (\ref{nonlinlag}) is
\be\label{lagphase}
L =  \left\{i\pi\cdot \dot\xi - \lambda^i\varphi_i\right\} + \mbox{c.c.}
\ee
where $\lambda^i$ are Lagrange multipliers for $n$ complex phase space
constraints, with constraint functions
\be\label{defvarphi}
\varphi_i = \pi_i - {\gamma \over2}\left[1 + \bar\xi\cdot\xi\right]^{-1}
\bar\xi_i\,.
\ee
Solving these constraints returns us to the original Lagrangian, up to a
total derivative. The $SU(n|1)$-supersymmetry transformations of the new
Lagrangian are
\bea\label{susyphase}
\delta \xi^i &=& \epsilon^i + \bar \epsilon\cdot \xi \xi^i\,, \nn
\delta \pi_i &=& {\gamma\over2}\bar\epsilon_i + \bar\epsilon_i\xi\cdot \pi -
\bar\epsilon\cdot\xi \pi_i\,, \nn
\delta \lambda^i &=& \bar\epsilon\cdot\xi\lambda^i +
\bar\epsilon\cdot\lambda\xi^i\,.
\eea

The $n$ complex constraint functions $\{\varphi\}$ are
equivalent to $2n$ real constraint functions that are second class, in
Dirac's terminology. However, the $n$ complex constraint functions are in
involution; it is only when we consider their complex conjugates that the
system of constraints becomes second class. In \cite{MR,HM} it was shown
that
when $2n$ real second class constraints can be separated into two sets of
$n$ real constraints in involution then one may quantize without
constraints on canonical space variables by imposing one set of $n$
constraints on the Hilbert space states and discarding the other
set\footnote{To our knowledge, for
systems with Grassmann second-class constraints the possibility of such a
quantization scheme was mentioned for the first time in \cite{CA}. The
idea behind it is that one set of constraints can be
interpreted as the $n$ gauge-fixing conditions for $n$ gauge-invariances
generated by the other set.}. Here we shall adopt an `analytic'
version of this procedure which, for Grassman variables, actually
preceded the formulation of the method for the real case; in this
context it has been called `Gupta-Bleuler quantization' \cite{LL,AL}.

As the method involves working with an unconstrained phase
space, the anticommutation relations follow directly from the canonical
Poisson brackets, and these may be realized by setting\footnote{The
classical anticommuting variable $\bar\pi$ is the complex
conjugate of the variable $\pi$, whereas the complex conjugate of
$\partial/\partial\xi$ is, for standard complex conjugation conventions,
$-\partial/\partial\bar \xi$. The conjugation in the quantum case
should be understood with respect to the properly defined
$SU(n|1)$ invariant scalar product, as discussed in section 4.}
\be
\pi_i = {\partial\over\partial\xi^i}\, ,\qquad
\bar\pi^i = {\partial\over\partial\bar \xi_i}\,.\label{conjugacy}
\ee
To take the constraints into account we require that physical states be
annihilated by the $n$ operators $\varphi_i$; this is equivalent to the
`analyticity' conditions\footnote{This is analogous to the chirality
condition on 4D chiral superfields, which arises in a similar way from
analytic quantization of the 4D superparticle \cite{Frydr,Dima}. See
\cite{bandos} for other analogous aspects of the superparticle case.}
\be
{\partial\Psi \over \partial \xi^i} = {\gamma\over2}\left[1 +
\bar\xi\cdot\xi\right]^{-1}\bar\xi_i\Psi\, ,
\qquad i=1,\dots,n\,,
\label{cond}
\ee
on wave-functions $\Psi(\{\xi\},\{\bar\xi\})$. These conditions have the
solution
\be
\label{solution2}
\Psi = \left[1+\bar\xi\cdot\xi\right]^{-{\gamma\over 2}} \Phi
\ee
for anti-analytic $\Phi$, which has the expansion
\be
\Phi = c_{(0)} + \bar\xi_i c_{(1)}^i  + \dots +
\bar\xi_{i_1}\dots
\bar\xi_{i_{n-1}}\ c_{(n-1)}^{i_1\dots i_{n-1}} + \bar\xi_{i_1}\dots
\bar\xi_{i_n}\ c_{(n)}^{i_1 \dots i_n}\,,\label{expansion}
\ee
where
\be
c_{(n-1)}^{i_1\dots i_{n-1}} \equiv {1\over (n-1)!}\varepsilon^{i_1\dots
i_{n-1}i_n}c_{(n-1)\,i_n}\,, \quad
c_{(n)}^{i_1 \dots i_n} \equiv {1\over n!}\varepsilon^{i_1\dots
i_n}c_{(n)}\ , \ldots~.
\ee

In principle each of the $2^n$ coefficients could have any Grassmann parity
but to implement $SO(2n)$ invariance we would have to choose all of them to
have the same Grassmann parity, which must be even for a positive definite
norm. In this way we would recover the standard free-fermion Hilbert space,
as a $2^n$-dimensional vector space, although the $SO(2n)$ invariance is not
manifest in our approach and has to be imposed. Here however, we wish to
explore the alternative possibility that the `Hilbert' space carries some
representation of the supergroup $SU(n|1)\,$. For this to be possible we
must take the anti-analytic function $\Phi$ to have a {\sl definite}
Grassmann parity\footnote{The same requirement is made in the standard
quantization of the superparticle, in contrast to the `spinning
particle'. In fact, as the 4-dimensional spinning particle and
superparticle Lagrangians can be shown to be {\sl classically}
equivalent \cite{VZ,STVZ,pkt}, the difference between the
two can be ascribed to
different quantization procedures, in close analogy to the
`free fermion' model considered here.}.
In this case the `Hilbert' space is a vector superspace of
dimension $(2^{n-1}|2^{n-1})\,$; for a reason to be made clear later, we
assume that $\Phi$ is Grassmann-even for $n$ even and Grassmann-odd for $n$
odd. Our next task is to determine how $SU(n|1)$ acts in this `Hilbert'
space.

\section{$SU(n|1)$ in `Hilbert' space}

The linear $U(n)$ transformations of the canonical variables of the
Lagrangian (\ref{lagphase}) are generated by the Noether charges
\be
\label{noether}
J^i{}_j = \bar\xi_j\bar\pi^i - \xi^i\pi_j\, .
\ee
The corresponding quantum $U(n)$ generators are the differential operators
\be
\hat J^i{}_j = \bar\xi_j{\partial\over\partial\bar\xi_i} -
\xi^i{\partial\over \partial\xi^j}\,.
\ee
For wave-functions of the form (\ref{solution2}) we have
\be
\hat J^i{}_j \Psi = \left[1 + \bar\xi\cdot\xi\right]^{-{\gamma\over2}}
\bar\xi_j{\partial\Phi\over\partial\bar\xi_i}\,,
\ee
from which we deduce the $U(n)$ transformation of $\Phi$ to be
\be
\delta_\omega\Phi
=\omega_i{}^j\, \bar\xi_j{\partial\Phi\over\partial\bar\xi_i}
\ee
where $\bar\omega^i{}_j = - \omega_j{}^i\,$.

The nonlinear supersymmetry transformations of (\ref{susyphase}) are
generated by the Grassmann-odd Noether charges
\be
S_i = \pi_i + {\gamma\over2}\bar\xi_i - \bar\xi_i\sum_j \bar\xi_j \bar\pi^j
\,
,\qquad  \bar S^i = \bar\pi^i + {\gamma\over2}\xi^i + \xi^i \sum_j
\xi^j\pi_j\,.\label{classS}
\ee
Note the presence of the terms linear in $\xi$ and $\bar \xi$; these
arise from the fact that the supersymmetry variation of the Lagrangian
is not zero but rather a total time derivative. These terms have no
effect on the transformations of $\xi^i$ generated by $\hat S$ and
$\hat {\bar S}$, which are those of (\ref{antransf}), but they {\it do}
contribute to the $U(1)$ charge in the $SU(n|1)$ superalgebra of Poisson
brackets of Noether charges. In fact, one finds that the $U(1)$ charge is
\be\label{bee}
B = \left({1\over n}-1\right) J^i{}_i + \gamma
\ee
where the shift by $\gamma$ is directly attributable to the
$\gamma$-dependent linear terms in $\hat S$ and $\hat {\bar S}$. In
passing to the quantum theory, the coefficients of these terms become
ambiguous because of operator ordering ambiguities. This ambiguity is
partially fixed by requiring that physical wave-functions $\Psi$ of the
form (\ref{solution2}) transform into physical wave-functions, i.e.
\be
\delta_\epsilon{\Psi}  \equiv -\left(\bar\epsilon\cdot \hat{\bar S} +
\epsilon\cdot S\right)\Psi =
\left[1+\bar\xi\cdot\xi\right]^{-{\gamma\over 2}}\delta_\epsilon{\Phi} .
\ee
This leaves us with the following quantum supersymmetry generators,
parametrized by a real number $\alpha$:
\bea
\hat S_i &=& {\partial\over\partial\xi^i} + {{\alpha}\over2}\bar\xi_i -
\bar\xi_i
\left(\bar\xi \cdot {\partial \over \partial\bar \xi}\right), \nn
\hat {\bar S}{}^i &=& {\partial\over\partial\bar\xi_i} +
{{\gamma}\over2}\xi^i +
\xi^i{\left(\xi\cdot {\partial \over \partial \xi}\right)}. \label{quantS}
\eea
These have the anticommutation relation
\be
\{\hat{S}_i, \hat{\bar S}^j \} = \left[\hat{J}^j{}_i -
{1\over n} \delta_i^j \hat J^k{}_k\right] + \delta_i^j \hat B
\ee
where $\hat B$ is the quantum $U(1)$ generator
\be
\hat B =  \left({1\over n}-1\right)\hat J^k{}_k + {1\over2}\left(\gamma +
\alpha\right).
\ee
One sees from this that the choice
\be
\alpha = \gamma
\ee
is `natural' because it leads to a quantum $U(1)\subset SU(n|1)$
generator that is the direct quantum counterpart of the classical
$U(1)$ charge $B$ of (\ref{bee}). Nevertheless, we shall consider the
case of general $\alpha$ in what follows.

We now compute the action of the charges $\hat S_i, {\hat{\bar S}}^i$, on
physical wave-functions. One finds that
\bea
\hat {\bar S}^i \Psi &=& \left[1+
\bar\xi\cdot\xi\right]^{-{\gamma\over2}}
\left[{\;\;\partial\Phi\over \partial
\bar\xi_i}\right] \nn
\hat S_i\Psi &=&  \left[1+
\bar\xi\cdot\xi\right]^{-{\gamma\over2}}
\left[{1\over2}{\left( \gamma + \alpha \right)\bar\xi_i }
-\bar\xi_i\left(\bar\xi \cdot
{\partial\over\partial\bar\xi}\right)\right]\Phi.
\eea
These results yield the following $SU(n|1)$-supersymmetry transformation
of $\Phi$:
\be\label{trans}
\delta_\epsilon\Phi =
-\left[q(\epsilon\cdot\bar\xi) + \bar\epsilon\cdot
{\partial\over\partial\bar\xi}
- \left(\epsilon\cdot
\bar\xi\right) \bar\xi\cdot
{\partial\over\partial\bar\xi}\right]\Phi
\ee
where
\be
q={1\over 2}(\gamma+\alpha)\,.
\ee
For component fields in the expansion (\ref{expansion}) this
transformation implies
\bea\label{transcomp}
&& \delta_{\epsilon} c_{(k)}^{i_1\ldots i_k} = (-1)^k\left\{ [k-1 -q]
\epsilon^{[i_1} c_{(k-1)}^{i_2 \ldots i_k]} -(1 -
\delta_{k,n})(k+1)\,\bar\epsilon_j c^{ji_1\ldots i_k}_{(k+1)}\right\}\;
(k \geq 2)\,,\nonumber \\
&& \delta_{\epsilon}c_{(0)} = - \bar\epsilon_i c_{(1)}^i\,, \quad
\delta_{\epsilon}c_{(1)}^i = q\epsilon^i c_{(0)} + 2\bar\epsilon_j
c^{ji}_{(2)}\;.
\eea
The full set of $SU(n|1)$ transformations of $\Phi$ are such that
\be\label{transphi}
\Phi'(\{\bar\xi'\}) =e^{iqs(\{\bar\xi\})} \Phi(\{\bar\xi\})
\label{gltran}
\ee
where $s$ is a local function of $\bar\xi$. Thus, $\Phi$ is a scalar
anti-analytic superfield when q=0; for other values of $q$, including
the `natural' value $q=\gamma$, one may consider $\Phi$ as a charged
scalar superfield, with charge $q$.

We have supposed up to now that $\alpha$ and $\gamma$ are arbitrary real
variables but one might expect the combination $q$ to be
quantized\footnote{For example, we have $q=\gamma$ for the `natural'
choice $\alpha=\gamma$ but, as mentioned earlier, $\gamma$ is the
coefficient of a WZ term. By analogy with the bosonic WZ terms, this
coefficient is expected to be quantized, though the origin of this
phenomenon can differ according to the case considered.}. As we
shall see, the representation
content of
the physical `Hilbert' space depends on $q$ and simplifications,
associated with the existence of zero norm states, occur for special
integer values of $q$.


\section{$SU(n|1)$-invariant norm}

In order to construct an invariant inner product one must first obtain an
$SU(n|1)$ invariant measure under the coordinate transformations
(\ref{antransf}). With the help of the lemma
\be
\delta_\epsilon \left[1+ \bar\xi\cdot\xi\right] =
\left(\bar\epsilon\cdot\xi -\epsilon\cdot\bar\xi\right)\left[1+
\bar\xi\cdot\xi\right],
\ee
it is not difficult to show that the $SU(n|1)$ invariant measure is such
that
\be
\int \!d\mu = \int \!d\mu_0 \left[1 + \bar\xi\cdot \xi\right]^{n-1}~,
\ee
where
\be
\int d\mu_0 = \prod_i {\partial\over\partial\bar\xi_i}
{\partial\over\partial\xi^i}\,.
\ee
However, because the transformation (\ref{trans}) involves a
$U(1)$ weight term, an additional factor is needed in the measure when
$q\ne0$. Let us replace $\Phi$ by $\Phi_{(q)}$ to remind us that $\Phi$
carries $U(1)$ charge $q$. Then the following bilinear form is
$SU(n|1)$ invariant:
\be
||\Phi_{(q)}||^2=
\int\! d\mu \left[1 + \bar\xi\cdot\xi\right]^{-q}
|\Phi_{(q)}|^2\,.
\label{norm2}
\ee
Note that the additional factor in the measure is unity precisely when
$q=0$ but is non-trivial otherwise.

In terms of the original wave-functions $\Psi =
\left[1+ \bar\xi\cdot \xi\right]^{-{\gamma\over 2}}\Phi_{(q)}$, the
$SU(n|1)$ invariant
scalar product corresponding to the definition (\ref{norm2}) reads
\be
<\Omega|\Psi> =\int\! d\mu_0 \left[1 + \bar\xi\cdot\xi\right]^{\kappa}
\Omega^{\dagger}\Psi \label{scalprod}
\ee
where
\be
\kappa = \gamma - q + n-1 \label{kappa}
\ee
and $\Omega(\xi, \bar\xi)$ is another vector in the same `Hilbert space'.
It is straightforward to check that the quantum generators (\ref{quantS})
are mutually conjugate with respect to this scalar product
\be
\left(<\Omega|\hat{S}_i|\Psi>\right)^\dagger = <\Psi|\hat{\bar S}{}^i
|\Omega >\,.
\label{selfS}
\ee
On the other hand, for $\kappa \neq 0$ the fermionic momentum operators
$\partial/\partial \xi^i$ and
$\partial/ \partial \bar\xi^i$ are not mutually conjugate with
respect to
(\ref{scalprod}). Note, however, that (\ref{scalprod}) is defined modulo
the following similarity transformation (change of basis) in
`Hilbert space'
\be
\Psi{} =\left[1+ \bar\xi\cdot \xi\right]^{\lambda }\Psi{}'_{(\lambda)}\,,
\quad
\Omega {} =\left[1+ \bar\xi\cdot \xi\right]^{\lambda }\Omega{}'_{(\lambda)}
\,,
\quad (\lambda^\dagger = \lambda)\,. \label{basises}
\ee
This amounts to the substitution  $\Omega, \Psi \rightarrow \Omega{}',
\Psi{}'$ and
shift $\kappa \rightarrow \kappa + 2 \lambda$ in the definition
(\ref{scalprod}), as well as a corresponding change in the observables.
The conjugacy property (\ref{selfS}) of the $SU(n|1)$ supersymmetry
generators is evidently basis-independent. In contrast, an analogous
conjugacy property holds for the fermionic momentum operators only for
the special choice of basis corresponding to $\lambda = - \kappa/2$:
\be
\left(<\Omega{}'_{(-\kappa/2)}|\partial/\partial
\xi^i|\Psi{}'_{(-\kappa/2)}>\right)^\dagger =
<\Psi{}'_{(-\kappa/2)}|\partial/\partial
\bar\xi_i|\Omega{}'_{(-\kappa/2)}>\,.
\label{selfpi}
\ee
Thus, the fermionic momentum operators (\ref{conjugacy}) are mutually
conjugate in the sense that there is a basis in `Hilbert space' for
which they satisfy (\ref{selfpi}).

Let us now turn to the analysis of the field content of $\Phi$ implied by
the invariant norm (\ref{norm2}).
In general, there are contributions to (\ref{norm2}) from all coefficients
in the expansion (\ref{expansion}), but zero norm states occur for special
values of $q$. For example, when $q=n-1$ we have
\be
||\Phi_{(n-1)}||^2 = |c_{(n)}|^2\,. \label{coeff22}
\ee
As $\delta_\epsilon c_{(n)} = 0$ for this choice of $q$ we see that the
physical Hilbert space is an $SU(n|1)$ singlet. All functions
$\Phi_{(n-1)}$ with $c_{(n)}=0$ have zero norm. If instead we set
$q= n-2$ then we find
\be
||\Phi_{(n-2)}||^2 = |c_n|^2 -{\bar c_{(n-1)}}^{\;\;i} c_{(n-1)\,i}\,.
\label{coeff}
\ee
Again there are zero norm states because the $SU(n)$ representation
content appearing in the norm is restricted to ${\bf n} \oplus {\bf 1}$;
these $SU(n)$ representations combine to form the fundamental ${\bf
n+1}$  representation of $SU(n|1)$. As a final example, consider $q=0$.
In this case we have
\be
||\Phi_{(0)}||^2 = (-1)^n (n-1)!
\sum^n_{k=1} (-1)^k \, k\,\, \bar c_{(k)\,i_1\ldots i_k}\,
c^{\;\;\;i_1\ldots i_k}_{(k)}\,. \label{generic}
\ee
The $SU(n)$ representation content is
${\bf n} \oplus {\bf n(n-1)}/2  \oplus \ldots \oplus {\bf n} \oplus {\bf
1}$.

An inspection of the transformations (\ref{transcomp}) leads to the
following general conclusions about the structure of the `Hilbert
spaces' corresponding to different choices of $q$. For the
choice
\be\label{condirr}
q= (\hat{k} -1)\,,
\ee
for integer $\hat{k}$ in the range $0 \leq \hat{k} \leq n$ (the
examples
above correspond to $\hat{k} = n,n-1,1$, respectively) there is an invariant
irreducible subspace spanned by
\be\label{set}
c_{(\hat{k})}^{i_1\ldots i_{\hat{k}}}\,, \dots\,, c^{i_1\ldots i_n}_{(n)}~.
\label{irr}
\ee
For $\hat{k}=0$ this subspace is the full space of coefficients of $\Phi$
but
otherwise it is not, and the remaining coefficients are transformed into
the above set; this shows that the representation of $SU(n|1)$ carried
by $\Phi_{(\hat{k}-1)}$ is reducible but not fully reducible. The norm
$||\Phi_{(\hat{k}-1)}||$ includes only the components (\ref{irr}), so there
exist zero norm states unless $\hat{k}=0$.

As the set (\ref{set}) is irreducible
under the action of $SU(n|1)$, we can consistently set them to zero:
\be
c_{(\hat{k})}^{i_1\ldots i_{\hat{k}}} = c_{(\hat{k} +1)}^{i_1\ldots
i_{\hat{k} +1}} =
\;\dots\; =
c^{i_1\ldots i_n}_{(n)} = 0~. \label{constrirr}
\ee
The complementary set of coefficients then forms an irreducible
set on its own, and one would expect there to exist a corresponding
$SU(n|1)$-invariant norm. However, the `obvious' norm, defined by
(\ref{norm2}), is identically zero when (\ref{constrirr}) is satisfied;
this is easily seen by rewriting (\ref{constrirr}) in the superfield form
\be
\label{superfconstr}
{\partial^{\hat{k}}\Phi_{(q)}\over \partial \bar\xi_{i_1}\ldots
\partial \bar\xi_{i_{\hat{k}}}} = 0\, \qquad (\mbox{and c.c.})\,.
\ee
One can check that these constraints are covariant under (\ref{trans})
provided that the condition (\ref{condirr}) holds. Of course, they do not
correspond to constraints in the Lagrangian (\ref{lagphase}) so what we
are now doing cannot be considered as a quantization of {\it that}
Lagrangian but one could add to it the classical
constraints corresponding to (\ref{superfconstr}), for which the constraint
functions are polynomials in $\bar\pi$.

It is remarkable that for $\Phi_{(\hat{k}-1)}$ constrained
by (\ref{superfconstr}) there exists the following {\it alternative} norm
\be\label{alternnorm}
|||\tilde{\Phi}_{(\hat{k}-1)}|||^2 = \int d\mu_0 \,(1 + \bar\xi\cdot\xi)^{n
-
\hat{k}}\,
\ln (1 + \bar\xi\cdot \xi)
\,\tilde{\bar{\Phi}}_{(\hat{k}-1)} \tilde{\Phi}_{(\hat{k}-1)}\,.
\ee
Taking into account that
\be
\delta_\epsilon \ln (1 + \bar\xi\cdot \xi) = \left(\bar\epsilon\cdot \xi -
\epsilon\cdot \bar\xi \right) \,,
\ee
it is straightforward to prove invariance of (\ref{alternnorm}) given
the constraints (\ref{superfconstr}), which are crucial to the result.
It is interesting that the `Lagrangian density' in (\ref{alternnorm}) is
not a tensor one as in (\ref{norm2}), but has an additional variation into
a total derivative, as is typical for WZ or Chern-Simons lagrangians.

\section{$n=2$ and BRST}

We shall now illustrate the above results with the $n=2$ case;
we also choose $\gamma=1$, which means that the `natural' choice of
operator ordering corresponds to $q=1$. For $n=2$ we can interpret
the odd coset space $SU(2|1)/U(2)$ as a BRST superspace because
$SU(2|1)\cong OSp(2|2)$ is the Euclidean BRST supergroup.  For $n=2$
we have
\be
\Phi_{(q)} = a + \bar\xi_i b^i + \bar\xi_1\bar\xi_2 c~.
\ee
The coefficients $(b^1,b^2)$, which form an $SU(2)$ doublet, can be
interpreted as (euclidean) Faddeev-Popov ghost and antighost for the
$SU(2)$-singlet gauge-fixing term $a$; the other $SU(2)$ singlet $c$
is then the `Nakanishi-Lautrup' auxiliary field.

>From (\ref{transcomp}) we deduce that the
$SU(2|1)$-supersymmetry transformations for $q=0$ are
\bea
\delta_\epsilon a &=& -\bar\epsilon_i b^i\,, \nn
\delta_\epsilon b^i &=& -\varepsilon^{ij}\bar\epsilon_j c\,,\nn
\delta_\epsilon c &=& \varepsilon_{ij} \epsilon^i b^j\,, \label{tr21}
\eea
where $\varepsilon_{12} = \varepsilon^{12} = 1, \;
\varepsilon^{ik}\varepsilon_{il} = \delta^k_l$.
This is not a reducible representation because $a$ transforms
non-trivially while $(b_1,b_2,c)$ span a 3-dimensional invariant subspace.
Observe that
\be
||\Phi_{(0)}||^2 = |c|^2  + b^i\bar b_i\label{norm21}
\ee
is invariant. Of course, this is not really a norm
as the variables $b^i$ are anticommuting.\footnote{For odd $n$ this
feature presents a difficulty because in this case the $c_{(n)}$ are
Grassmann odd and the $c^i_{(n-1)}$ are Grassmann even, but this
difficulty is overcome by changing the Grassmann parity of $\Phi$;
this is why we earlier required $\Psi$ and $\Phi$
to be even for $n= 2k$ and odd for $n=2k+1$. With this definition,
the norm for bosonic variables is always positive semi-definite.}
In other words, the physical states are vectors in a vector superspace
of dimension $(1|2)$ transforming as the fundamental representation of
$SU(2|1)\,$.

For $q=1$ the transformations (\ref{tr21}) become
\bea
\delta_\epsilon a &=& -\bar\epsilon_i b^i\,, \nn
\delta_\epsilon b^i &=& -\varepsilon^{ij}\bar\epsilon_j c + \epsilon^i
a\,,\nn
\delta_\epsilon c &=& 0\,.\label{tr22}
\eea
This is the `natural' case for which the physical Hilbert space is a
singlet. Indeed, the  norm (\ref{norm2}) in this case is simply
\be
||\Phi_{(1)}||^2 = |c|^2\,.
\ee

Still with $q=1$, we may impose the covariant condition
\be
c = 0 \quad \Leftrightarrow \quad
\frac{\;\;\partial^2 \tilde{\Phi}_{(1)}}{\partial \bar\xi_{i}\partial
\bar\xi_k} = 0~.
\ee
This leaves us with the irreducible multiplet $(a, b^i)$:
\bea
\delta_\epsilon a &=& -\bar\epsilon_i b^i\,, \nn
\delta_\epsilon b^i &=& \epsilon^i a\,.
\eea
The alternative norm is
\bea
|||\tilde{\Phi}_{(1)}|||^2 &=& - \int d\mu_0\,\ln (1 +\bar\xi\cdot \xi)\,
\tilde{\bar\Phi}_{(1)}\tilde\Phi_{(1)} \nn
&=& |a|^2 + \bar b_i b^i \label{norm3}
\eea
so physical states once again transform as a fundamental $({\bf 1} + {\bf
2})$ representation of $SU(2|1)$ (the precise correspondence with the
realization
(\ref{tr21}), (\ref{norm21}) is achieved via substitutions $a \rightarrow
\bar{\tilde{c}}$,
$b^i  \rightarrow \epsilon^{ik}\bar{\tilde{b}}_k$, where $\tilde{c}$ and
$\tilde{b^i}$
are transformed just as $c$ and $b^i$).

Finally, we shall consider $q=-1$, for which the transformation law
(\ref{trans}) becomes
\bea
\delta_\epsilon a &=& -\bar\epsilon_i b^i\,, \nn
\delta_\epsilon b^i &=& -\varepsilon^{ij}\bar\epsilon_j c - \epsilon^i
a\,,\nn
\delta_\epsilon c &=& 2\,\varepsilon_{ij} \epsilon^i b^j\, \label{tr23}
\eea
and the invariant norm calculated by the formula (\ref{norm2}) is
\be
||\Phi_{(-1)}||^2 = |c|^2 + 2|a|^2 - 2\bar b_ib^i~.
\ee
In this case one cannot single out any invariant subspace and
so ends up with a 4-dimensional irreducible multiplet $(b_1, b_2, c,
a)$ of $SU(2|1)$.
\section{Discussion}

We have shown that the mechanics of $n$ free complex Grassmann-odd variables
provides a non-linear realization of the supergroup $SU(n|1)$. It can be
viewed as the mechanics of a `particle' with the Grassmann-odd coset space
$SU(n|1)/U(n)$ as its phase space. This model is trivial in the sense that
its Hamiltonian vanishes but it is the simplest of a class of models that
realize $SU(n|1)$ non-linearly and for which the Hamiltonian is generically
non-zero. The particle on $SU(2|1)/[U(1)\times U(1)]$ is an example, and
one that will be considered in a future publication. Part of the
motivation of this paper was to exhibit some of the properties of these
models in the simplest possible setting. Another motivation is that
coset-spaces of the $n=2$ supergroup $SU(2|1)$ can be interpreted as
BRST superspaces.

We have shown that there exists an alternative quantization of the
`free fermion' model that implements the classical $SU(n|1)$ symmetry. In
contrast to the standard quantization, for which the states transform as a
spinor of $SO(2n)$, the states of the alternative quantum theory are vectors
in a vector superspace transforming under $SU(n|1)$. The specific
$SU(n|1)$ representation content depends on the resolution of an operator
ordering ambiguity, which amounts to a choice of $U(1)$ charge for the
wave-function. There is a natural choice, given the initial classical
Lagrangian, because this Lagrangian can be viewed as a WZ term for
$U(1)\subset SU(n|1)$ and this leads to specific shift in the $U(1)$
generator that is naturally identified with the $U(1)$ charge of the
quantum wave-function. For this choice, and a particular
choice of the `coupling constant',  the `Hilbert' space contains
zero norm states and the physical `Hilbert' space is an $SU(n|1)$
singlet.

For other choices of $U(1)$ charge assignment (and other choices of
coupling constant) one gets other representations of $SU(n|1)$,
picked out by an
$SU(n|1)$ invariant norm. We showed that there exists a class of integer
$U(1)$ charge assignments for which the representation is irreducible.
Remarkably, in this case the complementary representation contained in
the wave-function, again irreducible, could be picked out by a
different invariant, but not manifestly-invariant, norm provided that
the original representation was constrained to be absent; this case
corresponds to the quantization of the original free-fermion
Lagrangian with additional phase space constraints.

Any quantization of Grassmann-odd variables has to take into account
(explicitly or implicitly) second-class phase-space constraints. In our
case these were non-trivial because of a redefinition of variables needed
for analyticity of the $SU(n|1)$ transformations. We dealt with these
constraints by a variant of the `gauge unfixing' method involving a
separation of the constraints into analytic and anti-analytic
subsets in involution. It may be helpful if we sketch here how this
method can be used to covariantly quantize the massless 4D
superparticle, as done in \cite{Frydr,Dima}. The fermionic constraint
operators are the
supercovariant derivatives $D_\alpha$ and their complex conjugates $\bar
D_{\dot\alpha}$. These are not all second class (given $p^2=0$) because
the combinations $p^{\alpha\dot\alpha}\bar D_{\dot\alpha}$ and
$p^{\alpha\dot\alpha}D_\alpha$ are first class. Although we should
require that both of these first class operators annihilate physical
states $\Phi$ we need only impose $p^{\alpha\dot\alpha} D_{\alpha}\Phi=0$
explicitly if we also impose $\bar D_{\dot\alpha}\Phi=0$, as required for
`analytic quantization', because the other one is then implied. The
independent constraints are therefore $\bar D_{\dot\alpha}\Phi=0$ and
$p^{\alpha\dot\alpha} D_{\alpha}\Phi=0$ (because these imply $p^2=0$), but
these are just the free field equations for a massless chiral superfield.

\medskip
\section*{Acknowledgments}
\noindent
We thank S. Gukov, S. Krivonos,  M Berkooz, and L. Susskind, for  helpful
discussions.
L.M. thanks the theory group at JINR, the physics department
of the Weizmann Institute,  the GR group of DAMTP, and the TG of
Stanford University, for hospitality and partial financial support.
L.M. was supported in part by the National Science Foundation under grant
PHY-9870101. E.I. and A.P. acknowledge a partial support from INTAS grant
No. 00-0254. E.I. was partially supported by grants DFG No. 436 RUS
113/669, RFBR-DFG 02-02-04002 and RFBR-CNRS 01-02-22005.


\end{document}